\documentclass[11pt]{article}
\usepackage{moriond,epsfig}

\def\Journal#1#2#3#4{{#1} {\bf #2}, #3 (#4)}

\def\PRL{\em Phys. Rev. Lett.}
\def\PRD{{\em Phys. Rev.} D}

\def\be{\begin{equation}}
\def\ee{\end{equation}}
\def\bea{\begin{eqnarray}}
\def\eea{\end{eqnarray}}

\newcommand{\dzero}     {D\O}
\newcommand{\ttbar}     {\mbox{$t\bar{t}$}}

\newcommand{\met}       {\mbox{$\not\!\!E_T$}}
\newcommand{\rar}       {\rightarrow}

\begin{document}
\vspace*{4cm}
\title{TOP QUARK PRODUCTION CROSS SECTION AT THE TEVATRON}

\author{ E. SHABALINA \\ for the CDF and \dzero ~collaborations }

\address{University of Illinois at Chicago, Chicago IL, USA 60612}

\maketitle\abstracts{
An overview of the preliminary results of the top quark pair production 
cross section measurements    
at a center-of-mass energy of 1.96 TeV carried out by the CDF and \dzero 
~collaborations is presented. The data samples used for the analyses are 
collected in the current Tevatron run and correspond to an integrated
luminosity from 360 pb$^{-1}$ up to 760 pb$^{-1}$.    
}

\section{Introduction}
\label{intro}
The top quark was discovered \cite{discover} in 1995 at the Fermilab Tevatron
$p\bar{p}$ Collider at $\sqrt{s}=1.8$ TeV based on about 
50 pb$^{-1}$ of data per experiment. The increased 
luminosity and higher collision energy of $\sqrt{s}=1.96$ TeV of the current 
Tevatron run allow precise measurements of the top quark production and
decay properties.     
The latest theoretical calculations \cite{theory} for the $t\bar{t}$ production
cross section ($\sigma_{t\bar{t}}$) at NLO have an uncertainty ranging from 9\%
to 12\%. The precise measurement of $\sigma_{t\bar{t}}$
is not only of interest as a test of perturbative QCD, 
but it also permits to probe the effects of new physics. Such effects could 
lead to the $t\bar{t}$ cross section dependent on the final state of the top
quark pair. It is therefore necessary to measure $\sigma_{t\bar{t}}$ in all decay channels.    

In the Standard Model (SM), the top quark decays almost 100\% of the time to a $W$
boson and $b$-quark. Therefore, in $t\bar{t}$ events the final state
is completely determined by the $W$ boson decay modes. This paper covers 
the recent top quark pair cross section measurements performed by CDF 
and \dzero ~in the dilepton channels, where both $W$ bosons decay leptonically 
into an electron or a muon ($ee$, $e\mu$, $\mu\mu$), in the lepton + jets 
channels, where one of the $W$ bosons decays leptonically and the other one  
hadronically ($e$+jets, $\mu$+jets), and in the
all hadronic channel, where both $W$ bosons decay hadronically. 

\section{Dilepton channel}
In the detector a dileptonic final state is characterized by the presence of two 
isolated high $p_T$ 
leptons, two high $p_T$ $b$-jets and a large missing transverse energy ($\met$) 
from the two neutrinos. The background
contributions are pure instrumental effects, entirely estimated from data, and
irreducible physics backgrounds, derived from Monte Carlo simulations. 
Sources of the former are multijet, $W$+jets 
and $Z\rar \ell\ell$ events with mismeasured \met, misidentified  
jets or misidentified isolated electrons or muons, whereas the latter mainly
include $Z \rar \tau\tau$ where the $\tau$ leptons decay leptonically, 
and WW/WZ (diboson) processes.  

A summary of the expected \ttbar 
~signals, backgrounds and observed number of candidate events in a dataset corresponding to 
the integrated luminosity of 750 pb$^{-1}$ in the dilepton channels obtained by CDF is presented 
in Table \ref{tab:1}. The preliminary quoted \ttbar ~cross section for $m_{top}$ = 175 GeV yields          
$\sigma_{\ttbar}=8.3 \pm 1.5{\rm (stat)} \pm 1.0{\rm (syst)}\pm 0.5{\rm (lumi)}$ pb. 

\begin{table}[t]
\caption{Summary result in the dilepton channels obtained by CDF (in number of
events).}
\label{tab:1}       
\vspace{0.4cm}
\begin{center}
\begin{tabular}{|l|c|c|c||c|}
\hline
Source           & $ee$             &    $\mu\mu$      & $e\mu$   & $\ell\ell$            \\
\hline
Total background  & 6.10$\pm$1.87 & 7.52$\pm$2.12 & 5.72$\pm$1.38 & 19.34$\pm$4.26 \\
$\ttbar$ ($\sigma_{t\bar{t}}$ = 6.7 pb) & 8.25$\pm$0.38 & 8.57$\pm$0.39 & 19.27$\pm$0.88 & 36.09$\pm$1.24     \\
Total SM expectation & 14.35$\pm$2.08  & 16.09$\pm$2.30 & 24.99$\pm$1.75 & 55.43$\pm$5.11     \\\hline\hline
Candidates in 750 pb$^{-1}$ & 12               & 24               & 28             &64        \\
\hline
\end{tabular}
\end{center}
\end{table}

The inclusive analysis performed by CDF makes an attempt to fit several SM processes 
that constitute the dilepton sample taking advantage of 
their separation in the $\met$-$N_{jet}$ phase space. 
$\ttbar$ and WW events typically have large $\met$ from the final state neutrinos,
but $\ttbar$ events have more  
jet activity. Conversely, $Z \rar \tau\tau$ events have small $\met$ originating from 
leptonic decays of $\tau$'s, and low jet activity. 
Cross sections extracted from fitting of the two dimensional $\met$-$N_{jet}$ distribution from the 
data to those from the expected SM contributions are summarized in 
Table \ref{tab:2}. In 
the case of the $ee$ and $\mu\mu$ channels only $\ttbar$ and WW cross sections are fitted since the 
additional cut on the $\met$ significance applied in these channels to reduce 
large 
$Z \rar ee (\mu\mu)$ backgrounds makes it hard to extract the $Z \rar \tau\tau$ 
cross section. The measured 
cross sections are in good agreement with the SM predictions 
\footnote{$\sigma_{\ttbar}$ is given for $m_{top} = 178$ GeV.}.   

\begin{table}[h]
\caption{Summary of the cross sections from the inclusive dilepton analysis 
for 360 pb$^{-1}$ by CDF.}
\label{tab:2}       
\vspace{0.4cm}
\begin{center}
\begin{tabular}{|l|c|c|}
\hline
Cross section           & $e\mu$               & $\ell\ell$            \\
\hline
$\ttbar$ & $9.3^{+3.1}_{-2.6}(\rm {fit}) ^{+0.7}_{-0.2}(\rm {shape})$ pb & $8.4^{+2.5}_{-2.1}(\rm {fit}) ^{+0.7}_{-0.3}(\rm {shape})$ pb \\
WW       & $12.3^{+5.3}_{-4.4}(\rm {fit}) ^{+0.5}_{-0.1}(\rm {shape})$ pb & $16.1^{+5.0}_{-4.3}(\rm {fit}) ^{+0.8}_{-0.2}(\rm {shape})$ pb     \\
Z/DY $\rar \tau\tau$ & $292.7^{+48.9}_{-45.1}(\rm {fit}) ^{+5.9}_{-2.9}(\rm {shape})$ pb &  \\
\hline
\end{tabular}
\end{center}
\end{table}
  
The statistical sensitivity of the dilepton channel can be improved by loosening the lepton 
identification criteria and selecting events with one well identified high $p_T$ lepton 
and one high $p_T$ isolated track and considering events with at least one jet.
Additional discrimination of the $\ttbar$ signal from the backgrounds is achieved by using 
$b$-jet identification ($b$-tagging). 
To distinguish a heavy-flavor jet (arising from a $b$- or $c$-quark) from a
light-flavor jet ($u$-, $d$-, $s$-quark or gluon) one can make use of the
presence of charged tracks significantly displaced from the
primary vertex due to the finite lifetime of the $B$- or $D$-meson
(lifetime tagging).  
\dzero ~measured $\sigma_{\ttbar}$ using lepton plus track events with one or 
more jets tagged by 
a lifetime $b$-tagging algorithm and combined it with the cross 
section extracted from $e\mu$ events \footnote{$e\mu$ events with well
identified electron and muon were vetoed in the 
lepton plus track analysis.}. 
The combined cross section based on a 370 $pb^{-1}$ dataset 
yields $\sigma_{\ttbar}=8.6^{+1.9}_{-1.7}{\rm (stat)} \pm 1.1{\rm (syst)}\pm 0.6{\rm (lumi)}$ pb. 
 
\section{Lepton + jets channel}
The signature of the $\ell$+jets channel consists of one isolated high $p_T$ 
lepton, $\met$ due to the neutrino and at least four jets. The dominant 
background processes are $W$+jets and multijet production. 
To discriminate signal from background, which is significantly higher in $\ell$+jets
channels compared to the dilepton ones two approaches are used. The first 
approach makes use of the distinct kinematic features of a \ttbar ~event arising 
from its large mass. It combines kinematic event information into a discriminant
or artificial neural network (ANN) and performs a fit to the data.   
The second approach requires that at least one of the jets per event is 
identified as a $b$-jet. 
In both approaches, at the first stage of the analysis a data sample enriched 
in $W$+jets and \ttbar ~events is defined. The remaining QCD multijet 
background originates primarily from $\pi^0$'s and $\gamma$'s 
misidentified as jets ($e$+jets channel) or from heavy flavor decays 
($\mu$+jets channel), and is evaluated directly from data.

\begin{figure}
\hskip -1.5cm
\begin{tabular}{ccc}
\mbox{\includegraphics[width=0.35\textwidth,clip=]{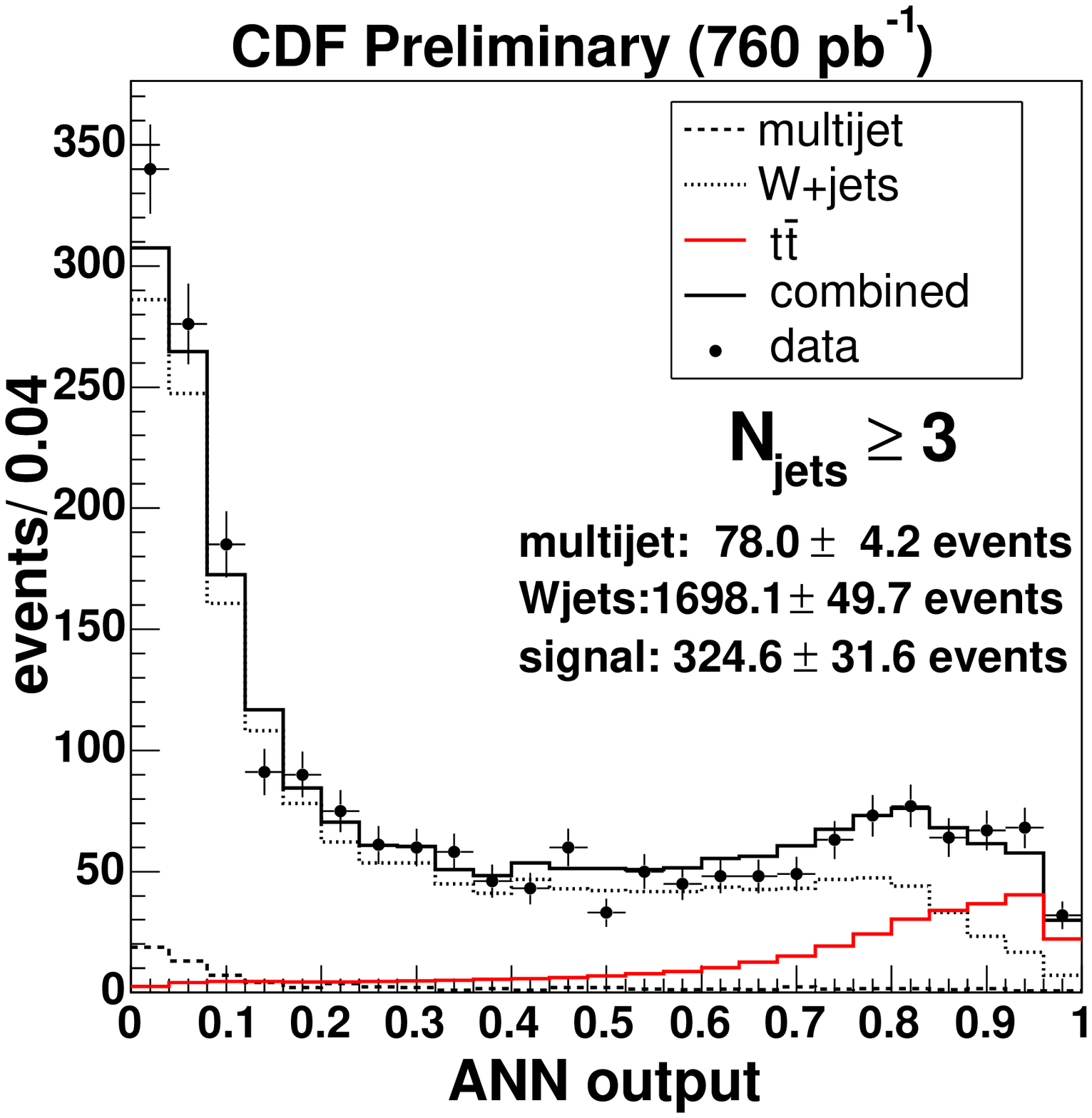}} &
\mbox{\includegraphics[width=0.35\textwidth,clip=]{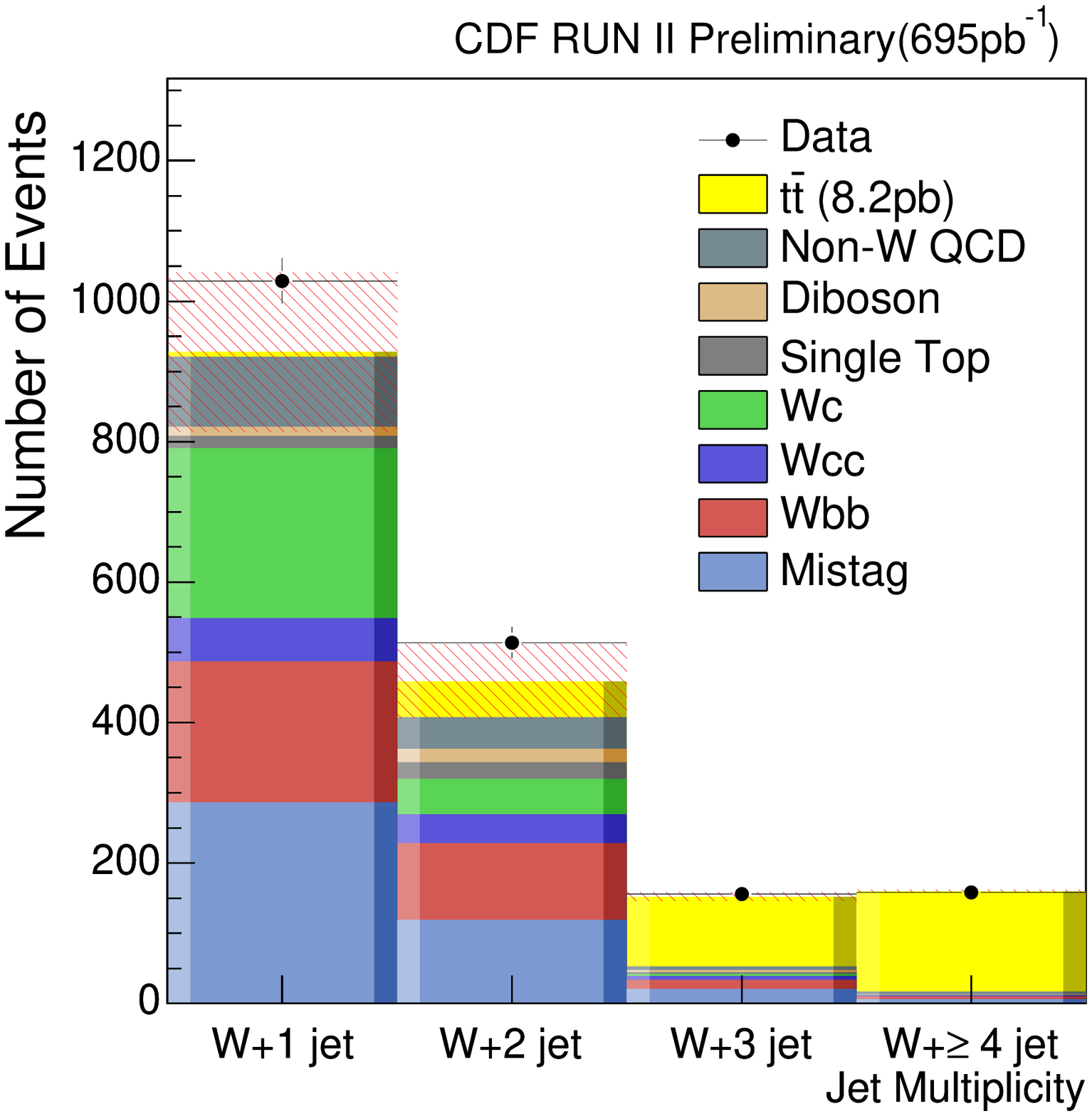}} &
\mbox{\includegraphics[width=0.35\textwidth,clip=]{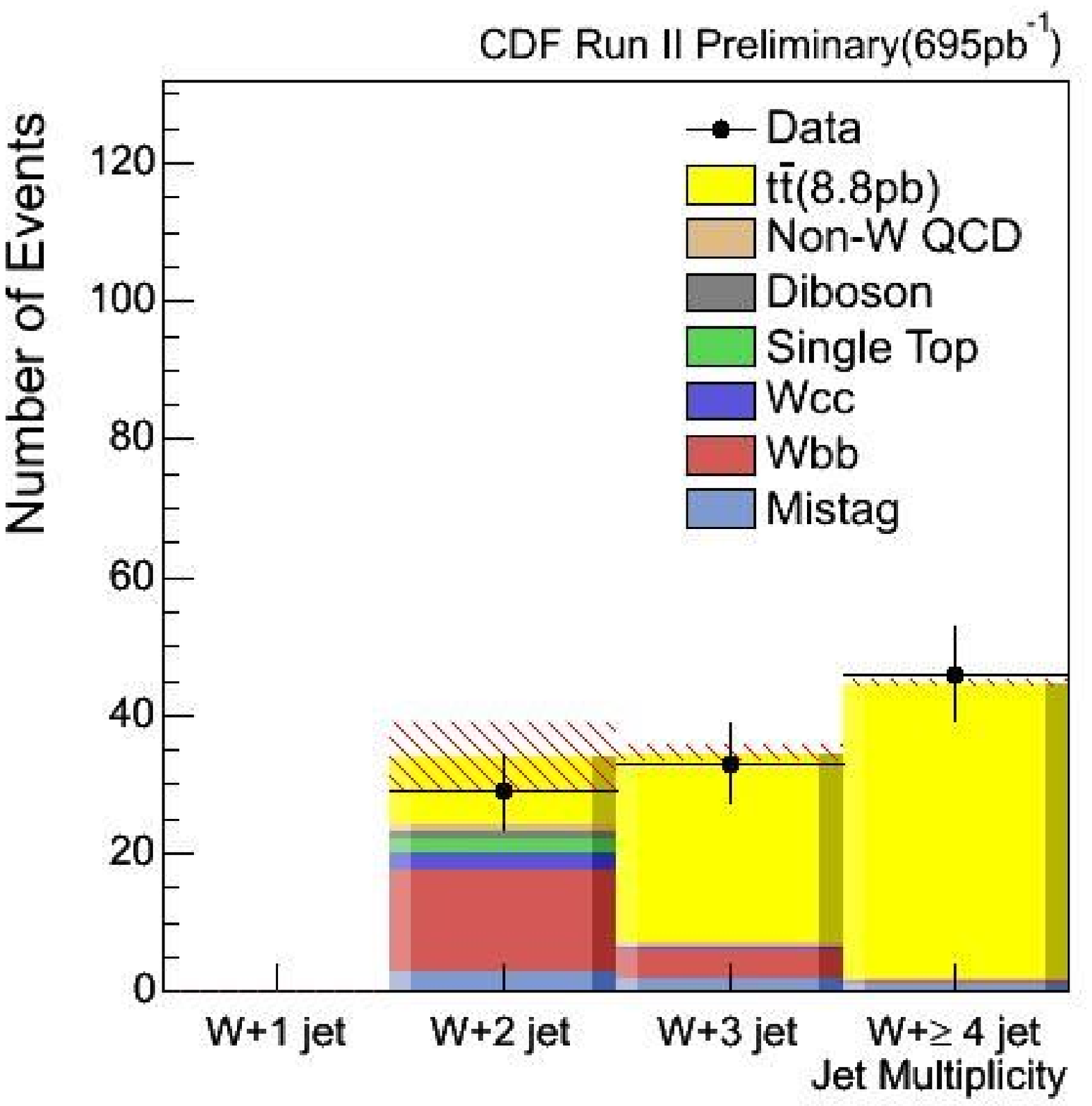}}
\put(-470,121){\bf{a}}\put(-300,121){\bf{b}}\put(-125,121){\bf{c}}
\end{tabular}
\caption{(a) Observed ANN output distribution versus fit result for 
$W$+$\ge$3 jet events; 
summary of backgrounds and measured \ttbar ~signal with at least 
one tag
(b) and $\ge 2$ tags (c) compared to the observed number of tagged events in data 
as a function of jet multiplicity. The band in (b) and (c) shows $\pm 1\sigma$ variation 
of the background.
\label{fig:cdf_topo}}
\end{figure}

CDF uses the ANN technique employing information
from seven input variables providing good discrimination between \ttbar
~signal and $W$+jets background. The results of the fit are shown in 
Fig.\ref{fig:cdf_topo}(a) for the events with 3 or more jets. The corresponding 
\ttbar ~cross section for a luminosity of 760 pb$^{-1}$ is  
$\sigma_{\ttbar}=6.0 \pm 0.6{\rm (stat)} \pm 0.9{\rm (syst)}$ pb.        

Both CDF and \dzero ~have measured $\sigma_{\ttbar}$ using 
lifetime tagging algorithms performing explicit reconstruction of 
secondary verteces with a large decay length significance with respect to the primary 
vertex. \dzero ~extracts the $t\bar{t}$ production cross section from the excess 
observed in the actual number of tagged events in $e$+jets and $\mu$+jets data with 
3 and $\geq 4$ jets and with exactly one and two or more tags with respect to the 
background expectation. 
Each source of systematic uncertainty included through a Gaussian term into the likelihood 
function used for the $\sigma_{\ttbar}$ calculation is allowed to affect the central value 
of the cross section, thus yielding a combined statistical and systematic uncertainty 
on $\sigma_{\ttbar}$.   
Assuming a top quark mass of 175 GeV \dzero ~measures 
$\sigma_{\ttbar}=8.1^{+1.3}_{-1.2}{\rm (stat+syst)} \pm 0.5 {\rm (lumi)} $ pb using 
a dataset of 360 pb$^{-1}$.   

In order to further reduce dominant backgrounds in the inclusive three jet sample with 
at least one $b$-tag, CDF requires that the scalar sum of the lepton $p_T$, jet $E_T$, 
and missing $E_T$, $H_T$, is larger than 200 GeV.
Cross section measurement using a dataset of ~700 pb$^{-1}$ yields 
$\sigma_{\ttbar}=8.2 \pm 0.6{\rm (stat)} \pm 1.0{\rm (syst)}$ pb, the most 
precise single measurement of $\sigma_{\ttbar}$ so far. The         
cross section extracted from the sample with $\ge$ 3 jets and at least two tags is
$\sigma_{\ttbar}=8.8^{+1.2}_{-1.1}{\rm (stat)} ^{+2.0}_{-1.3}{\rm (syst)}$ pb.  
Contributions of various backgrounds and measured \ttbar ~signal compared to 
the observed number of tagged 
events for different jet multiplicity bins are summarized in 
Figures \ref{fig:cdf_topo}(b,c).

\section{All hadronic channel}
\begin{figure}
\centering
\begin{tabular}{cc}
\psfig{figure=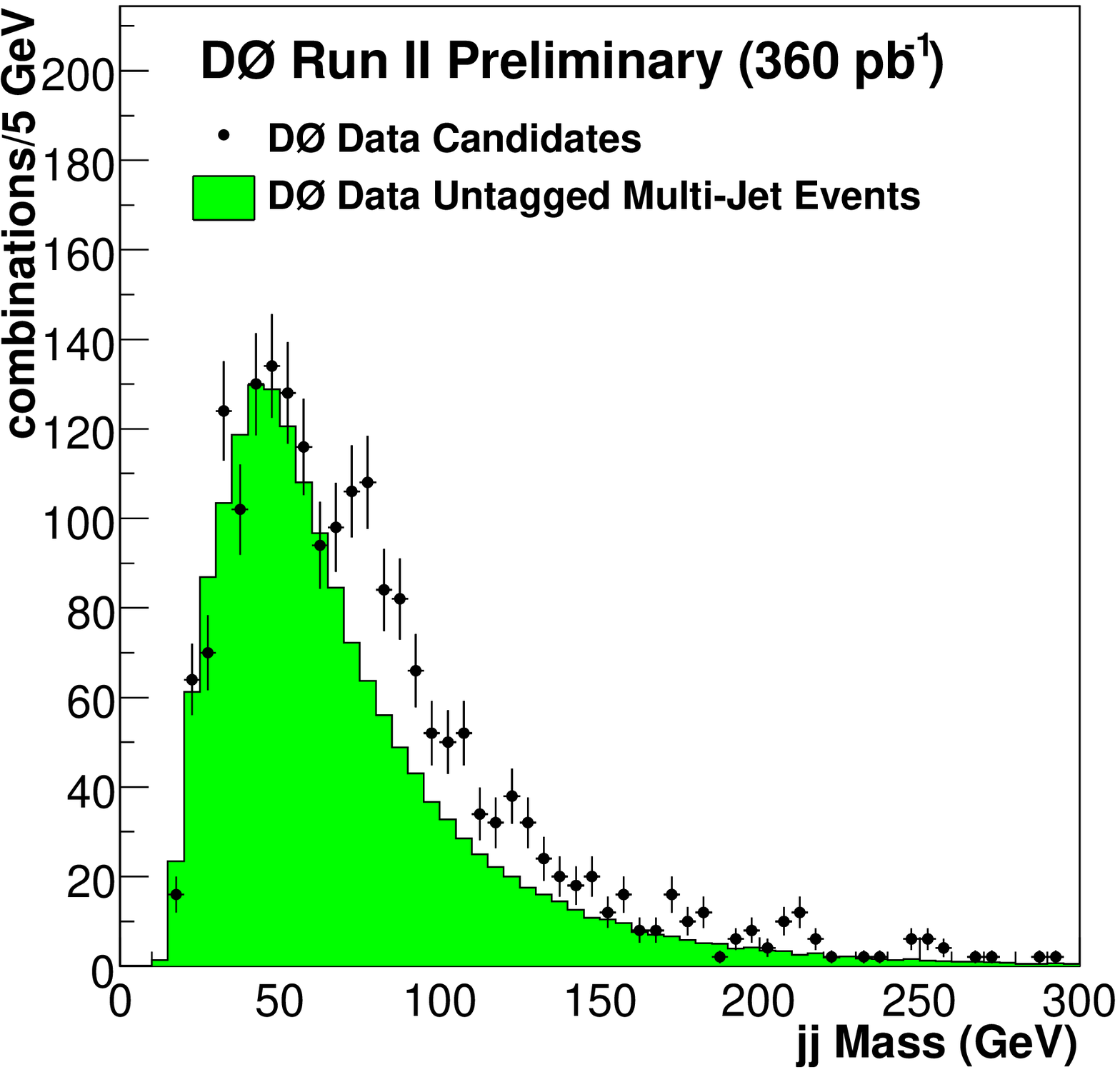,height=2.2in} &
\psfig{figure=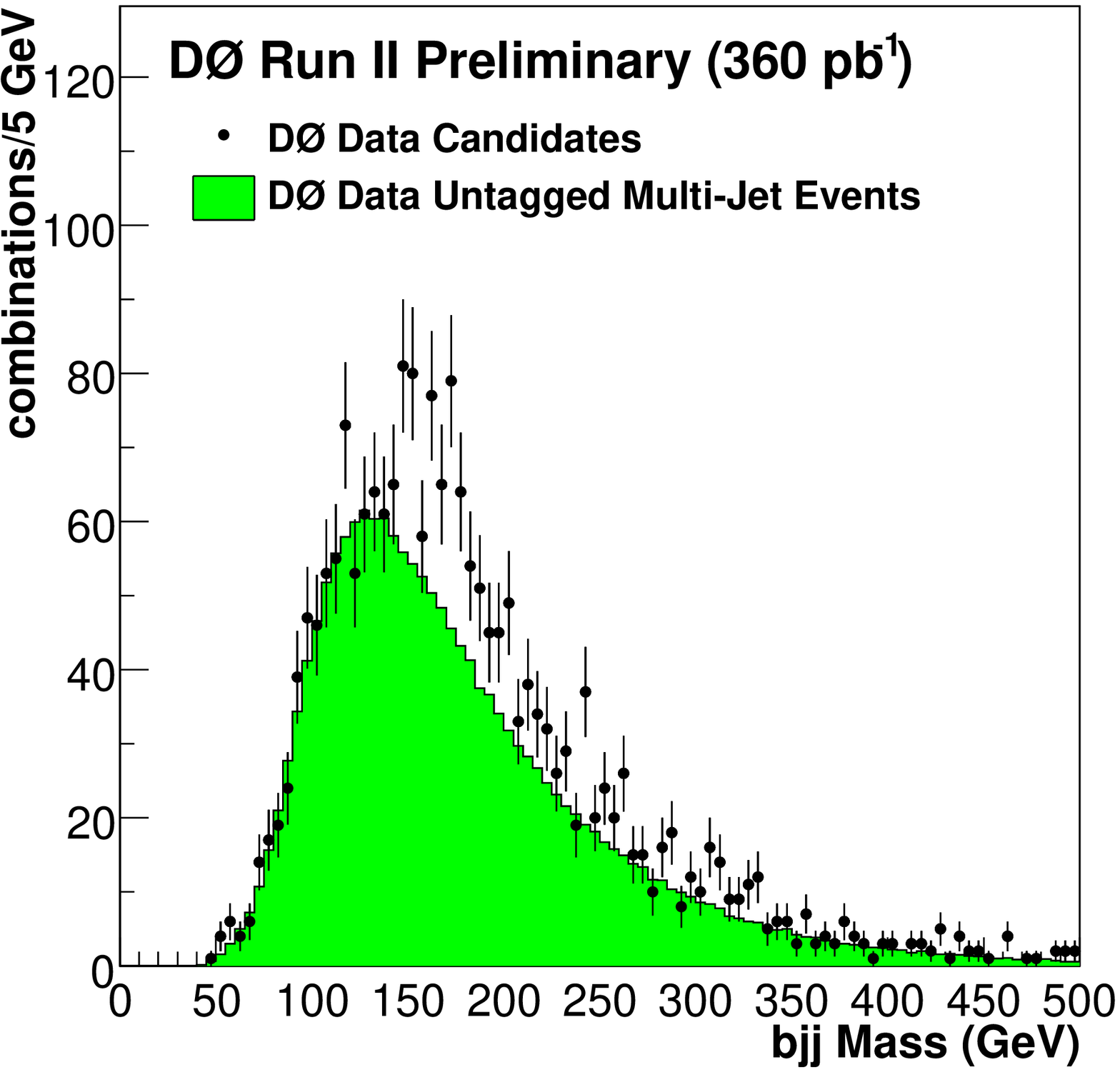,height=2.2in}
\end{tabular}
\caption{The dijet mass spectrum for all pairs of non $b$-tagged jets (left) 
and the three-jet mass spectrum for all combinations of one $b$-tagged jet and 
two non-$b$ tagged jets (right) with background overlayed.
\label{fig:alljets}}
\end{figure}

The all hadronic final state is characterized by six high $p_T$ jets 
two of which are $b$-jets. The dominant background in this channel, multijet 
production, is orders of magnitude larger than the \ttbar ~signal making 
the latter hard to identify. On the other
hand, the absence of the neutrinos in the final state allows to fully 
reconstruct the
\ttbar ~signal and thus discriminate it from the background. 
\dzero ~performs a $\sigma_{\ttbar}$ measurement by  
utilizing the three jet invariant mass distribution 
where one of the jets is identified as a $b$-jet and the other two are light jets for 
background normalization. 
Fig. \ref{fig:alljets} shows the dijet mass spectrum for all pairs of light jets 
revealing a $W$ boson peak (left) and the three jet mass spectrum (right). The 
shape of the overlaid background is determined from data by assigning $b$-flavor to a random jet. 
The excess of events over the background is attributed to the top quark production.
The cross section measured using a 360 pb$^{-1}$ dataset yields 
$\sigma_{\ttbar}=12.1 \pm 4.9{\rm (stat)} \pm 4.6{\rm (syst)}$ in agreement with
SM.  
     
\section{Conclusion}
Both CDF and \dzero have significantly improved the accuracy of 
the \ttbar ~cross section measurements in all decay channels using data collected 
in the current Tevatron run. CDF has combined six measurements of $\sigma_{\ttbar}$ 
achieving a 15\% improvement in the absolute uncertainty with
respect to the best single measurement. The combined result for 
$m_{top}$=175 GeV,  
$\sigma_{\ttbar}=7.3 \pm 0.5{\rm (stat)} \pm 0.6{\rm (syst)} \pm 0.4{\rm
(lumi)}$, is in good agreement with the theoretical prediction. Measurements in
all decay channels are consistent with each other and with the combined result.

\end{document}